# Incorporating Failure Knowledge into Design Decisions for IoT Systems: A Controlled Experiment on Novices

Dharun Anandayuvaraj, Pujita Thulluri, Justin Figueroa, Harshit Shandilya, James C. Davis
Purdue University, IN, USA
{ dananday, sthullur, figuero0, hshandil, davisjam }@purdue.edu

*Abstract*—Internet of Things (IoT) systems allow software to directly interact with the physical world. Recent IoT failures can be attributed to recurring software design flaws, suggesting IoT software engineers may not be learning from past failures. We examine the use of failure stories to improve IoT system designs.

We conducted an experiment to evaluate the influence of failure-related learning treatments on design decisions. Our experiment used a between-subjects comparison of novices (computer engineering students) completing a design questionnaire. There were three treatments: a control group (N=7); a group considering a set of design guidelines (N=8); and a group considering failure stories (proposed treatment, N=6). We measured their design decisions and their design rationales. All subjects made comparable decisions. Their rationales varied by treatment: subjects treated with guidelines and failure stories made greater use of criticality as a rationale, while subjects exposed to failure stories more frequently used safety as a rationale. Building on these findings, we suggest several research directions toward a failure-aware IoT engineering process.

## I. INTRODUCTION AND BACKGROUND

The Internet of Things (IoT), comprising smart devices interconnected with complex networks [1], has proliferated in modern societies [2]. In an IoT system, software interacts directly with the physical world, possibly autonomously, using distributed resources. A constellation of advances — in batteries, hardware, wireless networking, mobile computing, cloud services, and machine learning — has made widespread IoT systems feasible [3]. The worldwide IoT market is forecast to grow to an installed base of 30 billion devices by 2030 [4]. These trends have enabled IoT systems to become pervasive and increasingly interactive with the physical world where faults and defects can be safety critical. IoT systems are complex, with many opportunities for failure.

To increase the reliability of IoT systems, it is necessary to follow reliable design practices. Recent IoT failures [5] are a result of persistent single points of failure due to design flaws: lack of redundancy [6], [7], [8], [9], isolation [10], [11], [12], [13], [14], and authentication [15], [10], [11], [16], [17]. This pattern suggests that IoT engineers may not be learning from previous design failures [5]. This is problematic because IoT systems are often deployed in systems that are safety-critical, business-critical, and mission-critical [18].

Learning from engineering failures enables successful design [19], [20]. Historically, lessons from failures have influenced system design in engineering disciplines such as civil, mechanical, and aeronautical engineering [19]. With the proliferation of IoT systems, software systems are increasingly safety-critical; thus, we advocate for the practice of learning from system failures within software engineering. In the software engineering research literature, utilizing lessons from failures has been limited to postmortem practices for software *project* failures [21], [22], [23], [24]. Similarly, we motivate the opportunity for postmortem practices for software *system* failures. Specifically, we investigate the influence of system postmortems on design decisions.

To improve system design, we focus on design decisions and their rationales. For complex systems, design decisions greatly impact outcomes [25]. Design decision rationales are used to understand the justifications, alternatives considered, and trade-offs evaluated of design decisions [26]. Prior research to improve design decisions through rationale treatments have been limited to using general decision-making principles [27], reflective questions [28], and reminder cards with reflective questions [29]. Current materials to guide decision making are limited to guidelines, such as from government [30], industry [31], researchers [32], and textbooks [33]. Researchers have emphasized the importance of design decisions [34], and the consideration of their real world implications [35]. Thus we inquire: *Could we improve software design decisions using design failures as a learning treatment?*

We conducted an experiment to study the influence of failure stories on design decisions. Our subjects were novices (21 computer engineering students). We found that failure stories (proposed learning treatment) were just as effective as design guidelines (current practice) at enabling subjects to reason about the criticality of design decisions. However, failure stories had a greater effect at enabling subjects to reason about the safety implications of their design decisions.

## II. METHODOLOGY

### A. Problem Statement

Our goal is to examine the influence of failure stories on design decisions for IoT development:

**RQ:** How does the awareness of failed design stories influence *design decisions* and their *rationale*?

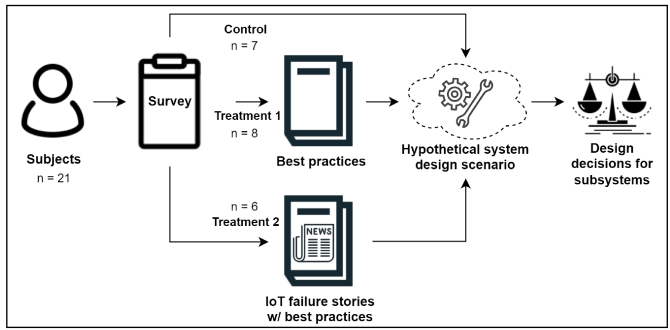

Fig. 1: Overview of experiment design.

### B. Study Design

*1) Compiling failure stories:* We selected 2 recent IoT failure incidents reported in news articles to compile our failure stories. The first story narrated a fatal plane crash due to a lack of redundancy in a critical IoT system [6]. The second story described smart-car hacks that allowed access to driving functions due to improper segmentation of the driving systems [12]. We identified the context, the cause, and the impact of the failures from the articles. We presented this information as a short paragraph in Treatment 2's questionnaire. We also extracted design practices to mitigate these failures. These design practices were presented in Treatment 2's questionnaire as "lessons" — a common postmortem practice [21]. These design practices were also presented as design guidelines in Treatment 1's questionnaire, but without the accompanying story. The stories are included in §VII.

To illustrate, we present one of the stories:

> **Story 1:** Recently, 2 airplanes crashed killing around 300 people. The cause of the crashes was identified as the interconnected stall protection system. This system was designed to stabilize the angle of an airplane from an unsafe upwards angle to a safe angle. This system consisted of one sensor to measure the angle, actuators to adjust the angle, a wireless network for communication, and a control system. The root cause of the failure was due to erroneous data from the sensor continuously triggering the system until the planes crashed. According to experts, a design redundancy of an additional sensor for this safety-critical system could have prevented the crashes.
> **Lesson 1**: Design redundancy for critical systems

*2) Simulated design scenario:* We created a hypothetical system design scenario to study the subjects' decision-making. The scenario depicted an IoT-enabled robotic e-commerce warehouse. We created 8 design decisions for subsystems in the warehouse. For each of the 2 failure stories, there were 4 design decisions that resembled the lesson outlined by the story, for a total of 8 design decisions. Also, 4 design decisions were for critical subsystems, whereas the other 4 design decisions were for non-critical subsystems. Each design decision consisted of two choices — a "correct" choice and an "incorrect" choice with respect to the criticality of the component and the lesson of a failure story. However, we note that these decisions are rarely binary, but rather ranked by criticality since any subsystem could be justified as critical. In order to account for this complexity, we also provided an open response field for the subjects to state their rationale for each of their decisions.

In addition, we included a budget for the overall system design scenario as a realistic constraint on design decisions [36]. The cost of each decision correlated with the criticality of the decision, and critical decisions necessitated a costlier choice.

To illustrate, we present one of the design decisions:

> **Design Decision 1:**
> Subsystem: Robot collision detection
> Subsystem Description: A system to detect and avoid an object in a robot's forward path.
> Design Question: The design team would like your help selecting the number of infrared sensors to use to detect objects in a robot's forward path.
> Choose between the two options for the missing component:
> Option 1: One infrared sensor (Cost: $10,000)
> Option 2: Two infrared sensors (Cost: $20,000)
> Please describe the reason for your decision.
>
> *The correct answer to this question is Option 2, since it is a safety-critical system resembling Story 1.*

*3) Study protocol:* The overview of the study protocol is outlined in Figure 1. This protocol was established after multiple rounds of pilot studies with 6 subjects to adjust the treatments and design scenarios. A budget constraint was added to the experiment as a result of pilot study feedback.[1] This study protocol was approved by our institution's IRB.

We recruited computer engineering students at our university through convenience sampling. Students were recruited at various stages of the undergraduate and graduate levels, inclusive of students with internship and full-time work experience. The questionnaire was distributed in a between-subjects design [37]. There were three treatment groups:

1) Control: Perform 8 subsystem design decisions.
2) Treatment 1 (current practice): Read 2 design guidelines (from the 2 failure stories) and perform 8 subsystem design decisions.
3) Treatment 2 (proposed): Read 2 failure stories + guidelines and perform 8 subsystem design decisions.

### C. Analysis

To determine whether the treatments influenced decisions, we studied the variance in decisions between the three groups. Since our sample size was limited, we utilized the Kruskal-Wallis Rank Sum test. We tested with a null hypothesis that the decisions are the same between the two conditions ($\alpha = 0.5$).

Since design decisions are complex, we also studied their rationales. With the qualitative data collected from the open response fields, we first performed open coding on a sample set of the data to establish a coding scheme [38]. This coding scheme was used to perform closed coding on all of the data [38]. Two authors conducted the qualitative analysis independently and the results were discussed to resolve conflicts.

---

[1]Subjects stated that without a financial constraint, they simply chose the most precautionary option for each decision.

## III. Result and Discussion

We received 21 responses to our questionnaire (Figure 2). Given our sample size, we frame our results as conjectures.

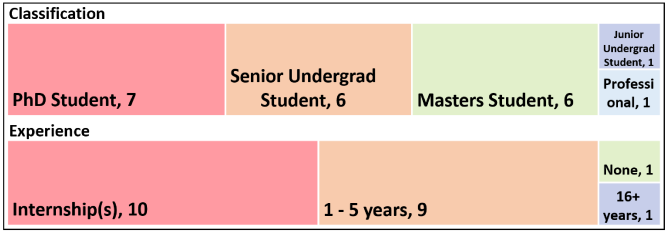

Fig. 2: The demographics of subjects in experiment.

We examined the influence of failure stories on IoT design decisions. Decisions do not vary by groups, neither visually (Figure 3) nor by Kruskal-Wallis Rank Sum test.

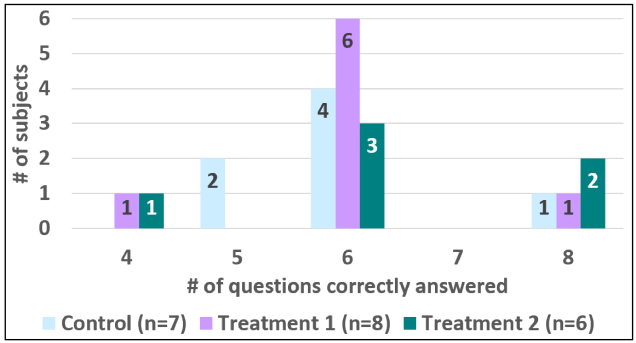

Fig. 3: Distribution of number of questions correctly answered.

With respect to decision rationales, we conducted a qualitative analysis. We coded each decision rationale using the codebook in Table I. The distribution of responses labeled by the codes is illustrated in Figure 4.

TABLE I: Coding scheme for decision rationales.

| Code | Definition |
| --- | --- |
| Criticality | Reasoned about the significance of decision |
| Safety | Reasoned about safety implications of decision |
| Cost | Used cost as a factor for decision |
| Performance | Used performance as a factor for decision |

We observed differences in decision rationales by group, notably in criticality and safety (Figure 4). The Treatment 1 and the Treatment 2 groups reasoned more about the criticality of the subsystems than the Control group did. The Treatment 2 group reasoned more about safety than the Control group, with Treatment 1 mentioning this least. The Control group was more concerned with cost and performance.

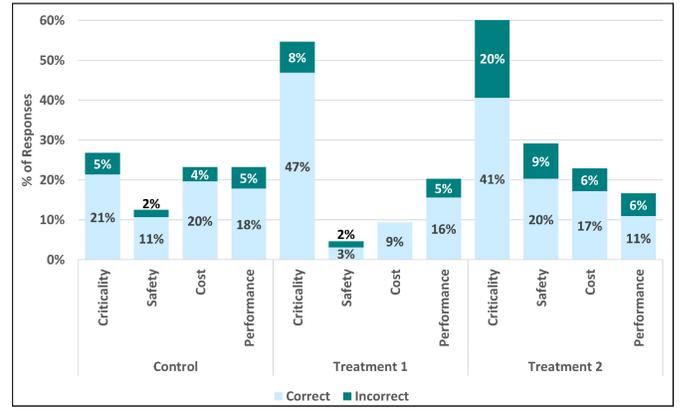

Fig. 4: Percent of responses by code, clustered by groups.

**Influence of failed design stories on design rationale concerning criticality.** As illustrated in Figure 4, we found that both treatments helped subjects reason about the criticality of their design decisions. Almost twice the amount of responses from Treatment 1 and Treatment 2 groups reasoned about criticality compared to responses from Control group. We conjecture that this might be due to the mention of criticality in the treatments. It is also noteworthy that a larger amount of responses *incorrectly* judge criticality from Treatment 2 than from Treatment 1 with respect to our assessment. We conjecture that this might be due to an over-precaution for criticality, due to the catastrophic impacts of bad design detailed in the stories, as a result of a priming effect [39].

To illustrate this trend in rationales concerning criticality, we provide example responses. For example, a question asked the subjects to make a decision about a crash alert system for robots with the option of only alerting the central warehouse management software or alerting the software as well as other nearby robots. We classified this subsystem as critical, necessitating redundancy with both alerts. A subject from Treatment 1 reasoned about criticality by identifying that *"robot crashes are critical."* Likewise, a subject from Treatment 2 said:

> *"In case a robot crashes, it would be important for other robots in its vicinity to know so they can avoid the crash. Otherwise, there could be more crashes..."*

In contrast, a Control subject was more concerned about cost:

> *"[It is] unlikely for multiple robots to...crash. Immediate human intervention can save $$."*

Thus, we conjecture that failure stories are effective at helping engineers reason about the criticality of design decisions.

**Influence of failed design stories on design rationale concerning safety.** As indicated in Figure 4, we found that Treatment 2 subjects more frequently reasoned about safety. It is also worth noting that the responses from the Control group considered safety more than the responses from Treatment 1. We conjecture this is due to the lack of constraints for the Control group subjects, enabling them to brainstorm factors (such as safety) to guide their decisions. This is consistent with prior findings [40].

To illustrate this trend in rationales concerning safety, we provide example responses. For example, a question asked subjects to make a decision about segmenting networks between personal and industrial devices, which could have safety implications under malicious circumstances. A Treatment 2 subject identified safety impacts of the decision:

> "If a worker clicks on malicious link or gets virus...on personal device, this could be dangerous if industrial controls are on same network. It is double the price, but safety should be higher concern."

A Control subject identified a malicious scenario — *"Attack on one can save the other"* — implying that they recognize the safety measure to isolate the critical subsystem. In contrast, the rationale of a Treatment 1 subject was limited to the design guidelines provided and did not consider safety:

> "Separation of core devices, Better security."

**Use of anecdotal logic.** Across all groups, 8 subjects used anecdotal logic as design rationales. For example, a Treatment 2 subject *correctly* cited the failure story we provided to choose redundancy for a safety-critical subsystem:

> "As we learned from the plane crash...design redundancy can help avoid terrible consequences. If the robots have only one sensor that becomes compromised, it could be hazardous for workers."

A Control subject *incorrectly* cited the university network as a basis to not segment personal and industrial networks:

> "I'm pretty sure all of the devices at [our university] share one or two connections so given that, it's probably good enough here as well."

A Treatment 1 subject likewise *incorrectly* cited the concept of robot swarms as a justification for maintaining a single communication channel for all robots in the warehouse:

> "I believe this is what swarm means, to have all the robots be controlled under a single connection."

Anecdotal logic is part of human nature [41]; if engineers use anecdotal logic, then failure stories provide relevant anecdotes.

## IV. FUTURE DIRECTIONS

Our initial results illustrate the potential of a failure-aware engineering process for developing IoT systems. We envision past failures informing various engineering phases including requirements elicitation, specification, design, implementation, validation, and maintenance. We outline a research agenda towards a failure-aware engineering process for IoT.

First, we propose an investigation into the effectiveness of guideline-based practices at enabling developers to reason about holistic (*e.g.,* safety, security, performance) implications of their design decisions. While we compared the influence of a guideline-based design practice against a failure-aware design practice, there is limited knowledge on the effectiveness of guideline-based design processes in the first place. As a step toward this insight, we plan a larger experiment using guideline-based treatments focused on safety-critical design decisions with constraints such as time, cost, and performance, again considering the design process and rationale.

Second, motivated by our difficulty studying design decisions at a binary level, we advocate for experimental methods to measure and understand design rationales. We suggest that qualitative analysis of rationales seems an appropriate path. A taxonomy of rationales could be a useful aid in experimental design. Measuring rationale in the context of the more systematic engineering techniques used in IoT design (*e.g.,* FMEA [42], STAMP [43]) is an open problem.

Lastly, we propose an investigation of processes for learning from design failures. If failure-based learning treatments are effective at instilling good design practices, then we need processes to identify failures and apply the knowledge. Software industry leaders advocate for improved postmortem practices [44], but lack empirical evidence. First, we need to investigate the current processes used by organizations. For example, practitioners could provide insight into whether and how organizations currently document and learn from system failures. Second, the knowledge transfer processes used to share system postmortems across teams and organizations could be studied. How effectively do teams generalize from a specific failure to the broader class? What are the limits of generalization? Third, we need to study the use of these failure stories in internal training (*e.g.,* during onboarding). When and why are failure stories shared to team members? What is learned, and how effective is it?

## V. THREATS TO VALIDITY

**Internal:** Our limited sample size and nonequivalent groups could have had confounding effects on experiment results [45]. Additionally, our experiment did not control for expertise.

**External:** Our sample size limits our findings to conjectures. Additionally, our questionnaire surveyed for intentions rather than actions, and did not incorporate a full design cycle. Furthermore, our subjects were students and their responses might not reflect industry practices.

**Construct:** We use design decisions and their rationale as proxy for design processes, which might not reflect industry practices.

## VI. CONCLUSION

We investigated the influence of failure stories on design decisions. We found that failure stories were as effective as design guidelines at guiding developers reason about the criticality of design decisions. We found that design guidelines constrained developers' ability to reason about the safety implications of design decisions, whereas failure stories were conducive. Our observations about failure-aware design decisions motivate new research directions into failure-aware design processes. We hope that this direction improves the safety of IoT systems.

## VII. DATA AVAILABILITY

Our questionnaire and data is available at https://doi.org/10.5281/zenodo.7724905. Our experiment protocol was approved by our institution's IRB (Purdue University IRB # 2022-200).